\documentclass[11pt]{article}

\usepackage{graphicx}
\usepackage{amsmath}
\usepackage{lineno}

\newcommand{\Fig}[1]{Figure~\ref{#1}}
\newcommand{\Equ}[1]{Eq.~(\ref{#1})}
\newcommand{\Sec}[1]{Section~\ref{#1}}
\newcommand{\Tab}[1]{Table~\ref{#1}}
\newcommand{\GEVcc}{\mbox{$\mathrm{GeV}/{{\it c}^2}$}}
\newcommand{\GEVc}{\mbox{$\mathrm{GeV}/{{\it c}}$}}

\newcommand{\MEVcc}{\mbox{$\mathrm{MeV}/{{\it c}^2}$}}

\newcommand{\PPI}{\mbox{$\pi^{+}$}}
\newcommand{\MPI}{\mbox{$\pi^{-}$}}
\newcommand{\PMP}{\mbox{$\pi^{\pm}$}}
\newcommand{\PIo}{\mbox{$\pi^{0}$}}

\newcommand{\KPL}{\mbox{$K^{+}$}}
\newcommand{\KMI}{\mbox{$K^{-}$}}
\newcommand{\KPM}{\mbox{$K^{\pm}$}}
\newcommand{\MEE}{\mbox{$m_{ee}$}}
\newcommand{\Pogg}{\mbox{$\pi^{0}_{\gamma\gamma}$}}
\newcommand{\Pod}{\mbox{$\pi^{0}_{D}$}}
\newcommand{\KPPEE}{\mbox{$\KPM \rightarrow \PMP \PIo e^{+}  e^{-}$}}
\newcommand{\PPEEP}{\mbox{$\KPL \rightarrow \PPI \PIo e^{+}  e^{-}$}}
\newcommand{\PPEEM}{\mbox{$\KMI \rightarrow \MPI \PIo e^{+}  e^{-}$}}
\newcommand{\KTAU}{\mbox{$\KPM \rightarrow \PMP \PPI \MPI$}}

\newcommand{\KTP}{\mbox{$\KPM \rightarrow \PMP \PIo$}}
\newcommand{\KPPG}{\mbox{$\KPM \rightarrow \PMP \PIo \gamma$}}

\newcommand{\PGG}{\mbox{$\PIo \rightarrow \gamma \gamma$}}
\newcommand{\KPPGS}{\mbox{$\KPM \rightarrow \pi^{\pm} \PIo \gamma^*$}}
\newcommand{\PPEE}{\mbox{$\PMP \PIo e^{+} e^{-}$}}
\newcommand{\PPD}{\mbox{$\PMP \pi^{0}_D$}}

\newcommand{\KDA}{\mbox{$\KPM \rightarrow \pi^{\pm} \PIo$}}
\newcommand{\PDAL}{\mbox{$\pi^{0}_D \rightarrow e^{+}  e^{-} \gamma$}}

\newcommand{\KPGS}{\mbox{$\KPM \rightarrow \pi^{\pm} \gamma^*$}}
\newcommand{\KSPGS}{\mbox{$K_{S} \rightarrow \PIo \gamma^*$}}

\begin{document}
\centerline{\LARGE EUROPEAN ORGANIZATION FOR NUCLEAR RESEARCH}
\vspace{15mm}
{\flushright{
CERN-EP-2018-246\\
September 5, 2018\\
 \vspace{2mm}
Revised version\\
November 22, 2018\\
 }}
\vspace{15mm}

\begin{center}
{\bf {\Large First Observation and Study of the {\boldmath $\KPPEE$} Decay \\
\vspace{5mm}
The NA48/2 Collaboration}}
\end{center}
\abstract{ 
The NA48/2 experiment at CERN reports 
the first observation  of the $\KPPEE$ decay from an exposure of $1.7 \times 10^{11}$ charged kaon decays recorded in 2003€--2004. A sample of 4919 candidates with  4.9\%  
background contamination allows the determination of the branching ratio in the full kinematic region,
$BR(K^{\pm} \to \pi^{\pm}\pi^{0}e^+e^-)=(4.24 \pm 0.14) \times 10^{-6}$. The study of the kinematic space shows evidence for a structure dependent contribution 
in agreement with predictions based on chiral perturbation theory.  
Several P- and CP-violating asymmetries are also evaluated.
}
\vspace{25mm}
\begin{center}
\em{Accepted for publication in Physics Letters B}
\end{center}
\clearpage
\begin{center}
{\Large The NA48/2 Collaboration}\\
\vspace{2mm}
 J.R.~Batley,
 G.~Kalmus,
 C.~Lazzeroni$\,$\footnotemark[1]$^,$\footnotemark[2],
 D.J.~Munday$\,$\footnotemark[1],
 M.W.~Slater$\,$\footnotemark[1],
 S.A.~Wotton \\
{\em \small Cavendish Laboratory, University of Cambridge,
Cambridge, CB3 0HE, UK$\,$\footnotemark[3]} \\[0.2cm]
 R.~Arcidiacono$\,$\footnotemark[4],
 G.~Bocquet,
 N.~Cabibbo$\,$\renewcommand{\thefootnote}{\fnsymbol{footnote}}%
\footnotemark[2]\renewcommand{\thefootnote}{\arabic{footnote}},
 A.~Ceccucci,
 D.~Cundy$\,$\footnotemark[5],
 V.~Falaleev$\,$\footnotemark[6], \\
 M.~Fidecaro,
 L.~Gatignon,
 A.~Gonidec,
 W.~Kubischta,
 A.~Maier,
 A.~Norton$\,$\footnotemark[7],\\
  M.~Patel$\,$\footnotemark[8],
 A.~Peters\\
{\em \small CERN, CH-1211 Gen\`eve 23, Switzerland} \\[0.2cm]
 S.~Balev\renewcommand{\thefootnote}{\fnsymbol{footnote}}
\footnotemark[2]\renewcommand{\thefootnote}{\arabic{footnote}},
 P.L.~Frabetti,
 E.~Gersabeck$\,$\footnotemark[9],
 E.~Goudzovski$\,$\footnotemark[1]$^,$\footnotemark[2]$^,$\footnotemark[10],
 P.~Hristov$\,$\footnotemark[11],
 V.~Kekelidze,
 V.~Kozhuharov$\,$\footnotemark[12]$^,$\footnotemark[13],
 L.~Litov$\,$\footnotemark[12],
 D.~Madigozhin,
  M.~Misheva$\,$\renewcommand{\thefootnote}{\fnsymbol{footnote}}%
\footnotemark[1]\renewcommand{\thefootnote}{\arabic{footnote}}$^,$\footnotemark[14],
  N.~Molokanova,
 I.~Polenkevich,
 Yu.~Potrebenikov,
 S.~Stoynev$\,$\footnotemark[15],
 A.~Zinchenko\renewcommand{\thefootnote}{\fnsymbol{footnote}}
\footnotemark[2]\renewcommand{\thefootnote}{\arabic{footnote}} \\
{\em \small Joint Institute for Nuclear Research, 141980 Dubna (MO), Russia} \\[0.2cm]
 E.~Monnier$\,$\footnotemark[16],
 E.~Swallow$\,$\renewcommand{\thefootnote}{\fnsymbol{footnote}}%
\footnotemark[2]\renewcommand{\thefootnote}{\arabic{footnote}},
 R.~Winston$\,$\footnotemark[17]\\
{\em \small The Enrico Fermi Institute, The University of Chicago,
Chicago, IL 60126, USA}\\[0.2cm]
 P.~Rubin$\,$\footnotemark[18],
 A.~Walker \\
{\em \small Department of Physics and Astronomy, University of Edinburgh, 
Edinburgh, EH9 3JZ, UK} \\[0.2cm]
 P.~Dalpiaz,
 C.~Damiani,
 M.~Fiorini,
 M.~Martini,
 F.~Petrucci,
 M.~Savri\'e,
 M.~Scarpa,
 H.~Wahl \\
{\em \small Dipartimento di Fisica e Scienze della Terra dell'Universit\`a e 
Sezione dell'INFN di Ferrara, \\
I-44122 Ferrara, Italy} \\[0.2cm]
 W.~Baldini,
 A.~Cotta Ramusino,
 A.~Gianoli\\
{\em \small Sezione dell'INFN di Ferrara,
I-44122 Ferrara, Italy} \\[0.2cm]
 M.~Calvetti,
 E.~Celeghini,
 E.~Iacopini,
 M.~Lenti,
 G.~Ruggiero$\,$\footnotemark[19] \\
{\em \small Dipartimento di Fisica dell'Universit\`a e Sezione
dell'INFN di Firenze,\\
 I-50125 Sesto Fiorentino, Italy} \\[0.2cm]
 A.~Bizzeti$\,$\footnotemark[20],
 M.~Veltri$\,$\footnotemark[21] \\
{\em \small Sezione dell'INFN di Firenze, I-50019 Sesto Fiorentino, Italy} \\[0.2cm]
 M.~Behler,
 K.~Eppard,
 M.~Hita-Hochgesand,
 K.~Kleinknecht,
 P.~Marouelli,
 L.~Masetti, \\
 U.~Moosbrugger,
 C.~Morales Morales, 
 B.~Renk,
 M.~Wache,
 R.~Wanke,
 A.~Winhart$\,$\footnotemark[1] \\
{\em \small Institut f\"ur Physik, Universit\"at Mainz, D-55099
 Mainz, Germany$\,$\footnotemark[22]} \\[0.2cm]
 D.~Coward$\,$\footnotemark[23],
 A.~Dabrowski$\,$\footnotemark[11],
 T.~Fonseca Martin,
 M.~Shieh,
 M.~Szleper$\,$\footnotemark[24],\\
 M.~Velasco,
 M.D.~Wood$\,$\footnotemark[23] \\
{\em \small Department of Physics and Astronomy, Northwestern
University, Evanston, IL 60208, USA}\\[0.2cm]
 G.~Anzivino,
 E.~Imbergamo,
 A.~Nappi$\,$\renewcommand{\thefootnote}{\fnsymbol{footnote}}%
\footnotemark[2]\renewcommand{\thefootnote}{\arabic{footnote}},
 M.~Piccini,
 M.~Raggi$\,$\footnotemark[25],
 M.~Valdata-Nappi \\
{\em \small Dipartimento di Fisica dell'Universit\`a e
Sezione dell'INFN di Perugia, \\
I-06100 Perugia, Italy} \\[0.2cm]
 P.~Cenci,
 M.~Pepe,
 M.C.~Petrucci \\
{\em \small Sezione dell'INFN di Perugia, I-06100 Perugia, Italy} \\[0.2cm]
 F.~Costantini,
 N.~Doble,
 L.~Fiorini$\,$\footnotemark[26],
 S.~Giudici,
 G.~Pierazzini$\,$\renewcommand{\thefootnote}{\fnsymbol{footnote}}%
\footnotemark[2]\renewcommand{\thefootnote}{\arabic{footnote}},
 M.~Sozzi,
 S.~Venditti  \\
{\em Dipartimento di Fisica dell'Universit\`a e Sezione dell'INFN di
Pisa, I-56100 Pisa, Italy} \\[0.2cm]
 G.~Collazuol$\,$\footnotemark[27],
 L.~DiLella$\,$\footnotemark[28],
 G.~Lamanna$\,$\footnotemark[28],
 I.~Mannelli,
 A.~Michetti \\
{\em Scuola Normale Superiore e Sezione dell'INFN di Pisa, I-56100
Pisa, Italy} \\[0.2cm]
 C.~Cerri,
 R.~Fantechi \\
{\em Sezione dell'INFN di Pisa, I-56100 Pisa, Italy} \\[0.2cm]
 B.~Bloch-Devaux\renewcommand{\thefootnote}{\fnsymbol{footnote}}
\footnotemark[1]\renewcommand{\thefootnote}{\arabic{footnote}}$^,$\footnotemark[29],
 C.~Cheshkov$\,$\footnotemark[30],
 J.B.~Ch\`eze,
 M.~De Beer,
 J.~Derr\'e,
 G.~Marel, \\
 E.~Mazzucato,
 B.~Peyaud,
 B.~Vallage \\
{\em \small DSM/IRFU -- CEA Saclay, F-91191 Gif-sur-Yvette, France} \\[0.2cm]
 M.~Holder,
 M.~Ziolkowski \\
{\em \small Fachbereich Physik, Universit\"at Siegen, D-57068
 Siegen, Germany$\,$\footnotemark[31]} \\[0.2cm]
 S.~Bifani$\,$\footnotemark[1],
 M.~Clemencic$\,$\footnotemark[11],
 S.~Goy Lopez$\,$\footnotemark[32] \\
{\em \small Dipartimento di Fisica dell'Universit\`a e
Sezione dell'INFN di Torino,\\ I-10125 Torino, Italy} \\[0.2cm]
 C.~Biino,
 N.~Cartiglia,
 F.~Marchetto \\
{\em \small Sezione dell'INFN di Torino, I-10125 Torino, Italy} \\[0.2cm]
 H.~Dibon,
 M.~Jeitler,
 M.~Markytan,
 I.~Mikulec,
 G.~Neuhofer,
 L.~Widhalm$\,$\renewcommand{\thefootnote}{\fnsymbol{footnote}}%
\footnotemark[2]\renewcommand{\thefootnote}{\arabic{footnote}} \\
{\em \small \"Osterreichische Akademie der Wissenschaften, Institut
f\"ur Hochenergiephysik,\\ A-10560 Wien, Austria$\,$\footnotemark[33]} \\[0.5cm]
\end{center}

\setcounter{footnote}{0}
\renewcommand{\thefootnote}{\fnsymbol{footnote}}
\footnotetext[1]{Corresponding authors, email:  brigitte.bloch-devaux@cern.ch, milena.misheva@cern.ch}
\footnotetext[2]{Deceased}
\renewcommand{\thefootnote}{\arabic{footnote}}
\footnotetext[1]{Now at: School of Physics and Astronomy, University of Birmingham,  Birmingham, B15 2TT, UK}
\footnotetext[2]{
Supported by a Royal Society University Research Fellowship
(UF100308, UF0758946)}
\footnotetext[3]{Funded by the UK Particle Physics and Astronomy Research Council, grant PPA/G/O/1999/00559}
\footnotetext[4]{Now at: Universit\`a degli Studi del Piemonte Orientale e Sezione 
dell'INFN di Torino, I-10125 Torino, Italy}
\footnotetext[5]{Now at: Istituto di Cosmogeofisica del CNR di Torino,
I-10133 Torino, Italy}
\footnotetext[6]{Now at: Joint Institute for Nuclear Research, 141980 Dubna (MO), Russia}
\footnotetext[7]{Now at: Dipartimento di Fisica e Scienze della Terra dell'Universit\`a e Sezione
dell'INFN di Ferrara, I-44122 Ferrara, Italy}
\footnotetext[8]{Now at: Department of Physics, Imperial College, London,
SW7 2BW, UK}
\footnotetext[9]{Now at: School of Physics and Astronomy, The University of Manchester, Manchester, M13 9PL, UK}
\footnotetext[10]{Supported by ERC Starting Grant 336581}
\footnotetext[11]{Now at: CERN, CH-1211 Gen\`eve 23, Switzerland}
\footnotetext[12]{Now at: Faculty of Physics, University of Sofia ``St. Kl.
Ohridski'', BG-1164 Sofia, Bulgaria, 
funded by the Bulgarian National Science Fund under contract DID02-22}
\footnotetext[13]{Also at: Laboratori Nazionali di Frascati, I-00044 Frascati, Italy}
\footnotetext[14]{Now at: Institute of Nuclear Research and Nuclear Energy of Bulgarian Academy of Science (INRNE-BAS), BG-1784 Sofia, Bulgaria}
\footnotetext[15]{Now at: Fermi National Accelerator Laboratory, Batavia, IL 60510, USA}
\footnotetext[16]{Now at: Centre de Physique des Particules de Marseille,
IN2P3-CNRS, Universit\'e de la M\'editerran\'ee, F-13288 Marseille,
France}
\footnotetext[17]{Now at: School of Natural Sciences, University of California, Merced, CA 95343, USA}
\footnotetext[18]{Now at: School of Physics, Astronomy and Computational Sciences, George Mason
University, Fairfax, VA 22030, USA}
\footnotetext[19]{Now at: Physics Department, University of Lancaster, Lancaster, LA1 4YW, UK} 
\footnotetext[20]{Also at Dipartimento di Scienze Fisiche, Informatiche e Matematiche, Universit\`a di Modena e Reggio Emilia, I-41125 Modena, Italy}
\footnotetext[21]{Also at Istituto di Fisica, Universit\`a di Urbino,
I-61029 Urbino, Italy}
\footnotetext[22]{Funded by the German Federal Minister for
Education and research under contract 05HK1UM1/1}
\footnotetext[23]{Now at: SLAC, Stanford University, Menlo Park, CA 94025, USA}
\footnotetext[24]{Now at: National Center for Nuclear Research, P-05-400 \'Swierk, Poland}
\footnotetext[25]{Now at: Universit\`a di Roma ``La Sapienza'', I-00185 Roma, Italy}
\footnotetext[26]{Now at: Instituto de F\'{\i}sica Corpuscular IFIC,
Universitat de Val\`{e}ncia, E-46071 Val\`{e}ncia, Spain}
\footnotetext[27]{Now at: Dipartimento di Fisica dell'Universit\`a e Sezione dell'INFN di Padova, I-35131 Padova, Italy}
\footnotetext[28]{Now at: Dipartimento di Fisica dell'Universit\`a e Sezione dell'INFN di Pisa, I-56100 Pisa, Italy}
\footnotetext[29]{Now at: Dipartimento di Fisica dell'Universit\`a di Torino, 
I-10125 Torino, Italy}
\footnotetext[30]{Now at: Institut de Physique Nucl\'eaire de Lyon,
IN2P3-CNRS, Universit\'e Lyon I, F-69622 Villeurbanne, France}
\footnotetext[31]{Funded by the German Federal Minister for Research
and Technology (BMBF) under contract 056SI74}
\footnotetext[32]{Now at: Centro de Investigaciones Energeticas
Medioambientales y Tecnologicas, E-28040 Madrid, Spain}
\footnotetext[33]{Funded by the Austrian Ministry for Traffic and
Research under the contract GZ 616.360/2-IV GZ 616.363/2-VIII, and
by the Fonds f\"ur Wissenschaft und Forschung FWF Nr.~P08929-PHY}


\clearpage

\section{Introduction and theoretical framework}\label{sec:intro}
Kaon decays  have played a major role in establishing the quark mixing flavour structure of the Standard 
Model \cite{Cirigliano:2011ny}. Radiative kaon decays  are of particular interest  in testing models
describing low-energy quantum chromodynamics (QCD) such as the chiral perturbation theory (ChPT), an effective field theory  valid below a scale $\cal O$(1 GeV). 

The radiative decay $\KPPEE$, never observed so far,  proceeds through virtual 
photon exchange followed by internal conversion into an electron-positron pair, 
i.e. $\KPPGS  \rightarrow  \PPEE$.
The virtual $\gamma^*$ can be produced by two different mechanisms: Inner Bremsstrahlung (IB)  where 
the $\gamma^*$ is emitted by one of the charged mesons in the initial or final state, and Direct Emission 
(DE) where the $\gamma^*$ is radiated off at the weak vertex. 
Consequently,  the  differential decay rate consists of three terms: the dominant long-distance IB contribution,
the DE component (electric E and magnetic M parts), and their interference. The interference term INT 
includes the different contributions, IB-E, IB-M and E-M.  The IB-M and E-M terms are P-violating and cancel upon angular integration in the total rate. 

There are few theoretical publications related to the $\KPPEE$  mode  
\cite{Pichl:2000ab,Cappiello:2011qc,Gevorkyan:2014waa} 
and no experimental observation. The authors 
of \cite{Cappiello:2011qc} predicted, on the basis of the NA48/2 measurement of the magnetic 
and electric terms involved in the $\KPPG$ decay \cite{Batley:2010aa}, the branching ratios of 
IB, DE and INT components of the $\KPPEE$ decay and posted recently a revised work 
\cite{neweval} where the interference term is re-evaluated using more realistic inputs based on 
additional experimental results and fewer theoretical assumptions.

It is worth writing explicitly the various contributions to the squared amplitude of the decay 
\cite{Cappiello:2011qc}:
\begin{equation}
\sum_{spins} |M|^2 = \frac{2e^2}{q^4} \left[ \sum_{i=1}^3 |F_i |^2 T_{ii} + 2 Re \sum_{i<j}^3 (F_{i}^{*} F_{j} )  T_{ij} \right],
\end{equation}\label{eq:msq}

\noindent  where $F_{i}$ are complex form factors and $T_{ij}$ are kinematic expressions (as defined in \cite{Cappiello:2011qc}) which depend on   
the four-momenta of the $e^+ e^-$ system and the  charged and neutral pions in the kaon rest frame. For convenience, one also writes:
\begin{equation}
F_1 = F_1 ^{IB} + F_1 ^{DE} , \phantom{xxxx}
F_2 = F_2 ^{IB} + F_2 ^{DE} , \phantom{xxxx}
F_3 = F_3 ^{DE}  .
\end{equation}
The form factors $F_1 ^{IB}, F_2 ^{IB}$  include a strong phase $\delta^2 _0$ corresponding to the S-wave  and isospin~2 state of the dipion system. 
The complex form factors $F_1 ^{DE}, F_2 ^{DE}$ correspond to the electric part of DE and make use of 
the ChPT  
counterterms $N_{E}^{(0,1,2)}$   while $F_3 ^{DE}$ corresponds to the magnetic part of DE and makes use of the counterterm  $N_{M}^{(0)}$. These form factors carry a strong phase $\delta^1_1$ corresponding to the P-wave  and isospin 1 state of the dipion system.

Numerical values of the counterterms were estimated \cite{neweval} using experimental measurements of form factors in the related modes $\KPGS$,  $\KSPGS$ and $\KPPG$.

\section{Kaon beam line and detector}\label{sec:beamdet}
The NA48/2 experiment at the CERN SPS was specifically designed for charge asymmetry measurements in the $\KPM$ $\to 3\pi$  decay modes \cite{Batley:2007}. Large samples of charged kaon  
decays were collected during the 2003--2004 data taking period.
The experiment beam line was designed to deliver simultaneous narrow momentum band $\KPL$ 
and $\KMI$ beams originating from  primary 400 GeV/$c$ protons extracted from the CERN SPS and 
impinging on a beryllium target. Secondary unseparated hadron beams with central momenta of 60~GeV/$c$ and a momentum band of $\pm$ 3.8\% (rms) were selected and brought to a common beam axis by two systems of dipole magnets with zero total deflection (called ``achromats''), focusing quadrupoles, muons sweepers and collimators.
The fraction of  beam kaons decaying in the 
114~m long cylindrical evacuated tank was 22\%.

The momenta of charged decay products were measured in a magnetic spectrometer, housed in a tank filled with helium at nearly atmospheric pressure. 
The spectrometer was composed of pairs of drift chambers (DCH) on each side of a dipole magnet  providing  a momentum kick $\Delta p = 120$ MeV/$c$  to charged particles in the horizontal plane. The momentum resolution achieved was $\sigma_p / p = (1.02 \oplus 0.044 \cdot p)$\% ($p$ in GeV/$c$).

A hodoscope (HOD) consisting of two planes of plastic scintillators, each segmented into 64 
strip-shaped counters, followed the spectrometer and provided time measurements for charged particles with a resolution of 150~ps.  Grouping the counters of each plane in eight subsets, the HOD surface was logically subdivided into 16 exclusive regions producing fast signals  used to trigger the detector readout on charged track topologies.

Further downstream was a liquid krypton electromagnetic calorimeter (LKr), an almost homogeneous ionization chamber with an active volume of 7~m$^3$, 
segmented transversally into 13248 projective $2\!\times\!2$~cm$^2$ cells with no longitudinal segmentation.  The energies of photons and electrons  were measured with  resolutions $\sigma_E / E= (3.2/\sqrt{E} \oplus  9.0/E \oplus 0.42)$\%.  The transverse positions of isolated showers were measured with a spatial resolution $\sigma_x = \sigma_y = (0.42/\sqrt{E} \oplus  0.06)$ cm, and the shower  time resolution was 2.5 ns $/\sqrt{E}$ ($E$ in GeV).
An iron/scintillator hadronic calorimeter and muon detectors were located further downstream. Neither of them was used in the present analysis. 

A dedicated two-level trigger was used to collect $\KPM$ decays into three charged tracks with  high efficiency: 
at the first level (L1), events containing charged tracks were selected by requiring space and time coincidences of signals in the two HOD planes in at least two of the 16 exclusive regions; at the second level (L2), a farm of asynchronous microprocessors performed a fast track reconstruction  and ran a 
vertex finding algorithm.

More details about the beam line and trigger implementation can be found in \cite{Batley:2007}.
A detailed description of the detector can be found in \cite{fa07}.

\section{Data analysis}\label{sec:datana}
\subsection{Measurement method}
The $\KPPEE$  decay rate 
is measured relative to the normalization decay  $\KTP$ collected concurrently with the same trigger logic.
This method does not rely on an absolute kaon flux measurement.  
In the signal sample, the $\PIo$  is identified through the $\PGG$  mode $(\Pogg)$. In the normalization sample, the $\PIo$  is identified through the $\PDAL$  Dalitz mode 
$(\Pod) $.
 The ratio of partial rates (and branching ratios) is obtained as:
\begin{equation}
BR(\KPPEE)/ BR(\KTP)= \frac{N_s -N_{bs}}{N_n - N_{bn}} \cdot \frac{A_n \times \varepsilon_n}{A_s \times \varepsilon_s} \cdot \frac{\Gamma(\Pod)}{\Gamma(\Pogg)},
\end{equation}
where $N_s , N_n$ are the numbers of signal and normalization candidates;  $N_{bs} , N_{bn}$ are the 
numbers of background events in the signal and normalization samples; $A_s$ and $\varepsilon_s$
are the acceptance and the trigger efficiency for the signal sample;  $A_n$ and $\varepsilon_n$
are those for the normalization sample.

The branching ratio of the normalization mode  is
$BR (\KTP)= (20.67 \pm 0.08) \%$  
and the ratio of $\PIo$ partial rates  is
 $\Gamma ( \Pod) / \Gamma ( \Pogg)= (1.188 \pm 0.035)\% $  \cite{pdg}.  Acceptances are obtained from a detailed Monte Carlo (MC) simulation based on GEANT3 \cite{GEANT}. The simulation includes full detector geometry and material description, stray magnetic fields,  DCH local inefficiencies  and misalignment, LKr local inefficiencies, accurate simulation of the kaon beam line and variations of the above throughout the data-taking period.

Efficiencies of the L1 and L2 triggers are measured from downscaled control samples, recorded concurrently with the three-track trigger. 
The control trigger condition for the L1 efficiency measurement requires  at least one coincidence of signals in the two planes of the HOD. The control trigger sample for the L2 efficiency measurement consists of L1 triggers recorded regardless of the L2 decision. The trigger decision is also available in the simulation for comparison.
\subsection{Event reconstruction and selection}
The standard NA48/2 software has been used including charged track, LKr energy cluster and 
 three-track decay vertex  reconstruction \cite{Batley:2007}.
Fully reconstructed $\KTAU$ decays have been used to monitor the DCH alignment,
the spectrometer field integral and the mean beam position at each DCH  plane throughout the data taking. 

Signal  and normalization candidates are reconstructed from three tracks: two same-sign tracks 
and one opposite-charge track forming a common vertex in the fiducial decay volume, the vertex charge 
being therefore  $q_{vtx} = \pm 1$.  The vertex time is defined as the average of the three HOD signal times  associated to the tracks. The tracks are required to be in time within 5 ns of the vertex time. Their impact points are required to be within the geometrical acceptance of the drift chambers.  In particular, the track distance to the monitored beam position in DCH1  plane is required to be larger than 12 cm.   The track momenta are required to be in the range (2--60) $\GEVc$ and track-to-track distances at DCH1 to be larger than 2~cm to suppress photon conversions to $e^+ e^-$ pairs in the upstream material.  

Configurations where the three considered tracks,  extrapolated 
to the HOD front face, have their impact points  in a single trigger region are rejected to avoid L1 inefficiencies of purely geometrical origin. Because of the different kinematics, this affects 2.3\% of the signal sample and has a negligible effect on the normalization sample. 

 All vertices considered for further analysis are required to be reconstructed in a 98 m long fiducial volume, starting 2 m downstream of the last collimator exit, and within 3~cm from the beam  axis.

Photon clusters matching  the vertex time within $5$ ns  are considered as  photon candidates if their  energy  is in  the range (3--60) GeV, their position is within the LKr geometrical acceptance and their distance to the nearest LKr inactive cell is larger than 2 cm. Photon four-momenta are reconstructed  assuming they originate from the three-track vertex. Photon trajectories are required to intercept the DCH1 plane at a radial position  larger than 11 cm to avoid possible interactions with the DCH flange resulting in a degraded energy measurement.
 
Signal and normalization modes differ in their final state by one photon, while satisfying similar kinematic constraints on the 
reconstructed $\pi^0$ and kaon masses, although with different resolutions because of different numbers of participating particles. The mass resolutions  (Gaussian rms) obtained from the data agree with those from simulation and are found  to be $\sigma_m (\Pod) \simeq 1.7~\MEVcc$, 
 $\sigma_m (\PMP \Pod )\simeq 4.2~\MEVcc$  and $\sigma_m (\Pogg) \simeq 2.7~\MEVcc$,  $\sigma_m (\PMP \Pogg e e) \simeq 6.1~\MEVcc$ for the normalization and signal modes, respectively.
 
 Very loose requirements are applied to the reconstructed masses, required to be  within 15~$\MEVcc$  (45~$\MEVcc$) from the nominal  $\pi^0$ ($\KPM$) mass \cite{pdg}, respectively, ensuring a minimal dependence of the selection on momentum or energy calibration effects, as well as on any resolution mismatches between data and simulation.   A common constraint, taking into account the correlation between the reconstructed m$_{\pi^0}$ and m$_K$ masses and  defined as
 \begin{equation}\label{eq:correl}
   | ~m_{\pi^0} - 0.42 \cdot m_{K} + 72.3  ~| < 6    {\rm  ~(all ~masses ~in~} \MEVcc),
 \end {equation}
   contains more than 99\% of the normalization events and about 96.5\% of the signal events.
   
 In both modes, the single track with its charge opposite to $q_{vtx}$ is considered to be an electron (positron). The remaining 
 $e / \pi$ ambiguity  for the two same-sign tracks is then solved by testing the two mass hypotheses against the full selection. 
 When a particular mass assignment is considered,  an extra requirement on the distance of any photon cluster to the track impact at the LKr front face is applied to guarantee photon shower isolation, avoiding potential overlap with other showers: 
 the distance between the photon position and the electron and positron track impacts  is required to  be larger than 10 cm and the distance between the photon position and the pion track impact to be larger than 20 cm. This requirement is enforced only for track impacts within the LKr geometrical acceptance.

No upper limit on the number of tracks and clusters is set, all three-track vertices being considered and combined with any photon cluster under the two possible $e / \pi$  mass hypotheses until one combination satisfies either of the following selections (normalization or signal) below, the event being rejected otherwise.
If both mass combinations are accepted, the one with the tighter constraint of  \Equ{eq:correl} is kept.
 \paragraph{Normalization selection}
The $\Pod$ candidate is reconstructed from  a pair of electron and positron tracks and a photon originating from  the three-track vertex.
The kaon candidate is reconstructed from the $\PPD$ system.

The consistency of the final state with a kaon decay along the beam axis is checked further by considering the energy-weighted coordinates of the centre of gravity (COG) of the particles at the LKr front plane computed from the photon position  and the track extrapolations obtained from track parameters measured before the magnet (undeviated trajectories).  The radial distance of the COG to the nominal beam position is required to be smaller than 2~cm.
The pion momentum is required to be larger than 10~$\GEVc$ and the total momentum of the system  to be in the beam momentum range (54--66)~$\GEVc$.
The  $e^+ e^-$ mass is required to be larger than 10~$\MEVcc$ to ensure good agreement between data and simulation. A sample of 16316690 candidates satisfies the normalization selection criteria. 
\paragraph{Signal selection}\label{sec:signal}
The $\Pogg$ candidate is reconstructed from  two photons originating from the three-track vertex. The kaon candidate is reconstructed from the $\pi^\pm \pi^0  e^+ e^-$ system. The two photon clusters are required to be separated by more than 10~cm at the LKr front plane to avoid shower overlap.
The event COG coordinates  are then obtained including the two photons and the three charged tracks, and subjected to the same requirement as above. 
 The total momentum of the system is  required to be in the beam momentum range (54--66)~$\GEVc$.
The  $e^+ e^-$ mass is required to be larger than 3~$\MEVcc$.

Two main sources of background  contribute to the signal final state: 
$K^{\pm} \to \pi^{\pm}\Pogg\pi^{0}_{D}$  ($K_{3\pi D}$) where one of the photons is lost 
(or merged with another particle), and $K^{\pm} \to \pi^{\pm}\pi^{0}_{D}(\gamma)$ ($K_{2\pi D\gamma}$),  
where the radiative photon and the  Dalitz decay photon mimic a $\pi^0 \to \gamma\gamma$ decay. 
Suppression of the $K_{3\pi D}$ background events is achieved by 
requiring the squared mass of the $\pi^+\pi^0$ system 
to be greater than 0.12~$(\GEVcc)^2$, exploiting the larger phase space available in the signal mode. 
This cut alone rejects 94\% of the $K_{3\pi D}$  simulated events and  $\sim$1\%  of the IB signal.
To reject  the $K_{2\pi D\gamma}$ background, each of the two possible masses $m_{ee\gamma}$ is required to be more than 7~$\MEVcc$ away from the nominal $\PIo$ mass
(corresponding to about $4 \sigma$ of the mass resolution). A sample of 4919 candidates satisfies the signal selection criteria.
\subsection{Background evaluation}
The background processes contributing to the normalization mode ($K_{2\pi  D}$) are  semi-leptonic decays followed by a Dalitz decay of the $\PIo$:  $K^\pm \to \mu^\pm  \nu \pi^0 _{D}$   ($K_{\mu3D}$) and  $K^\pm \to e^\pm \nu \pi^0 _D$ ($K_{e3D}$), collectively  denoted $K_{l3D}$, 
 where the $\pi^{0}_{D}$ decay is correctly reconstructed but the lepton ($\mu^\pm , e^\pm $) is erroneously attributed the $\PPI$  mass.   
The acceptances of such processes in the normalization selection are $\cal O$($10^{-4}$)  and obtained from large simulated samples.

For each background process, the number of  events  $N_{bn}$ is estimated relative to the number of observed events in the normalization mode $N_{n}$  using the acceptances in the  normalization selection and the world average branching ratios \cite{pdg}:
\begin{equation}\label{eq:bkgn}
  K_{l3D} : N_{bn} / N_n  =  (A_{K_{l3D }} / A_n) \cdot BR(K_{l3D}) /  BR(K_{2\pi  D})
\end{equation}
where the trigger efficiencies cancel to first order due to the similar topologies. 

The number of background events in the signal selection $N_{bs}$ is estimated relative to the number of observed events  in the normalization selection $N_{n}$ and is obtained as in \Equ{eq:bkgn}, using the acceptances in the signal selection, both $\cal O$($10^{-6}$):
\begin{equation}\label{eq:bkg1}
 K_{3\pi D} : N_{bs} / N_n = 2 \times  (A_{K_{3\pi D}} / A_n ) \cdot BR(K_{3\pi D}) \times BR(\PIo \to \gamma\gamma) / BR(K_{2\pi  D}),
 \end{equation}
\begin{equation}\label{eq:bkg2}
K_{2\pi D\gamma} : N_{bs} / N_n = A_{K_{2\pi D\gamma}} / A_n  .
\end{equation}
Note the factor of two in \Equ{eq:bkg1} due to the two $\PIo$ mesons  in the $K_{3\pi D}$ mode.
An order of magnitude  smaller contribution from $K_{e3D}$   is also considered.
In all contributions
both background and normalization branching ratios include the $\PIo$ Dalitz decay  partial rate whose value and uncertainty cancel in the estimation. 
\section{Branching ratio measurement}\label{sec:br}
\begin{figure}[htb]
\begin{minipage}{0.5\linewidth}
\includegraphics[width=1.\linewidth]{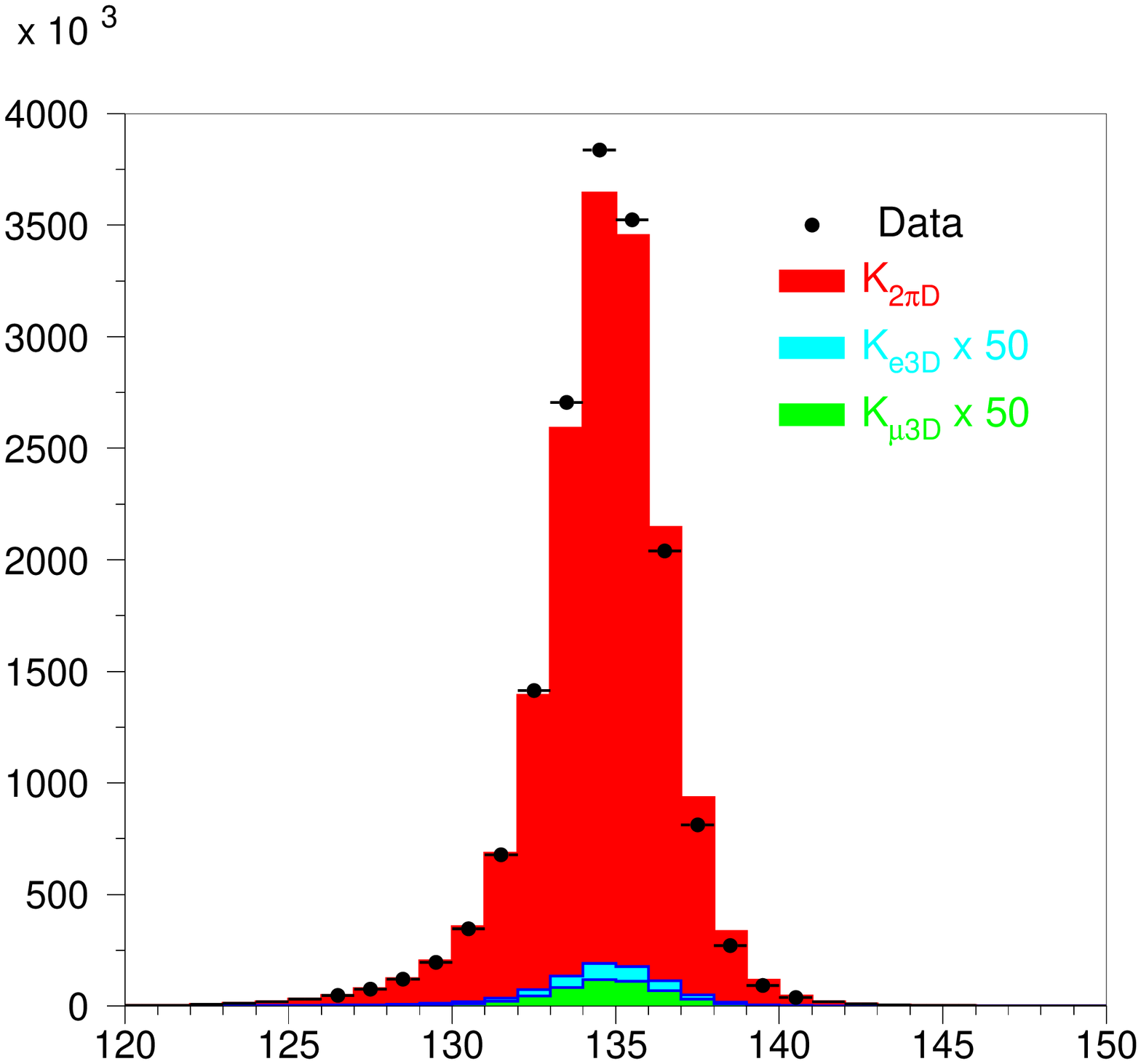}
\put(-94,-5){$m_{\gamma e^+ e^-}$ [$\MEVcc$]}
\put(-240,90){\rotatebox{90}{Events / (1 $\MEVcc$)}}
\end{minipage} 
\begin{minipage}{0.5\linewidth}
\includegraphics[width=1.\linewidth]{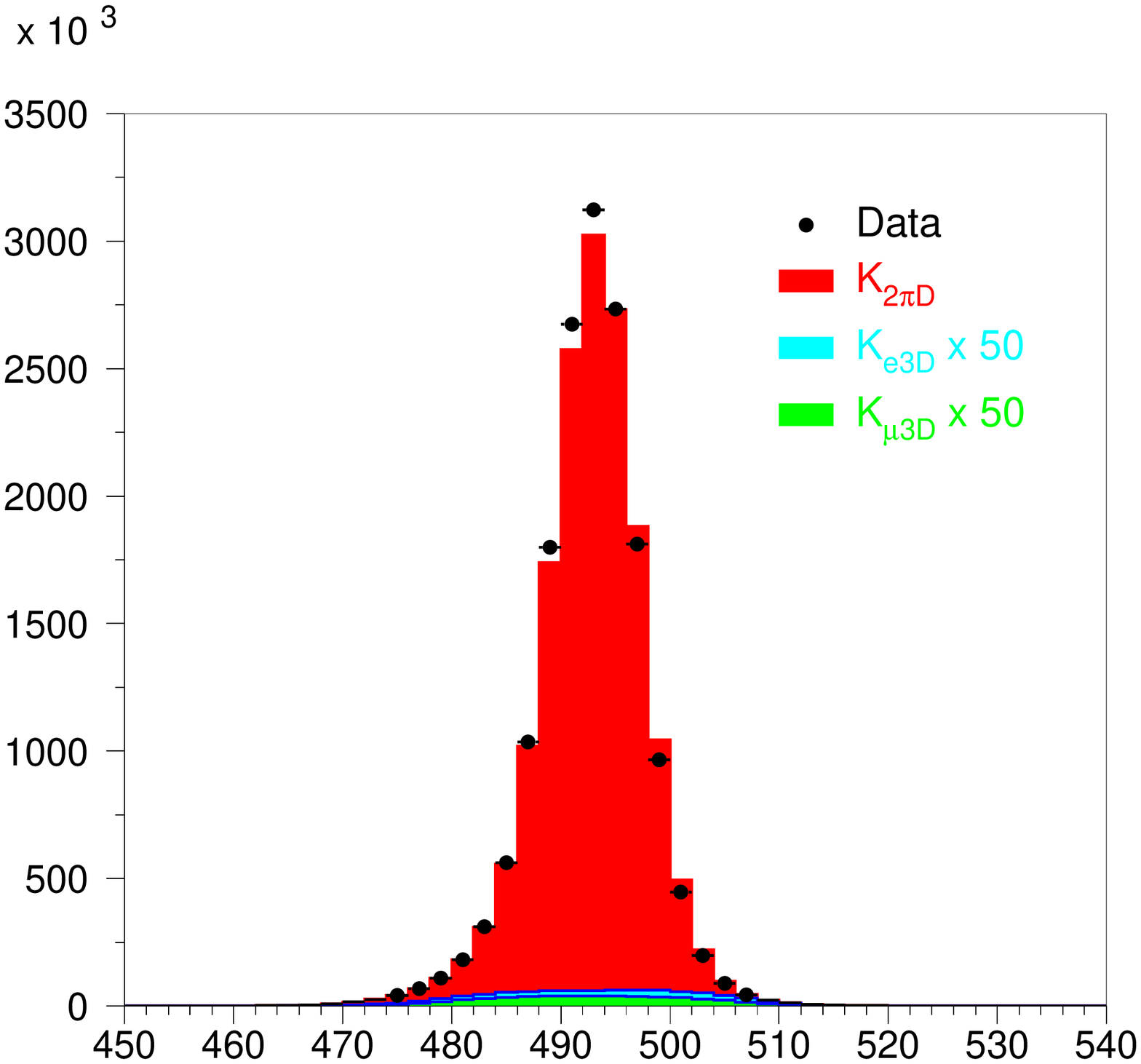}
\put(-90,-5){$m_{\pi \pi^0 _D}$ [$\MEVcc$]}
\put(-240,90){\rotatebox{90}{Events / (2 $\MEVcc$)}}
\end{minipage}  
\caption{ \label{fig:pp0d} Normalization candidates. Left: reconstructed $\gamma e^+ e^-$ mass. Right: reconstructed  $\pi^\pm  \pi^0 _D$ mass.  Full dots correspond to data candidates; stacked histograms are, from bottom to top, the expected $K_{\mu3D}$ (green) and $K_{e3 D}$ (blue) backgrounds multiplied by a factor of 50 to be visible.  The normalization  simulation (red) includes radiative effects in both kaon and $\pi^{0}_D$ decays that reproduce the asymmetric tails of both distributions. }
\end{figure}

\begin{figure}[htb]
\begin{minipage}{0.5\linewidth}
\includegraphics[width=1.\linewidth]{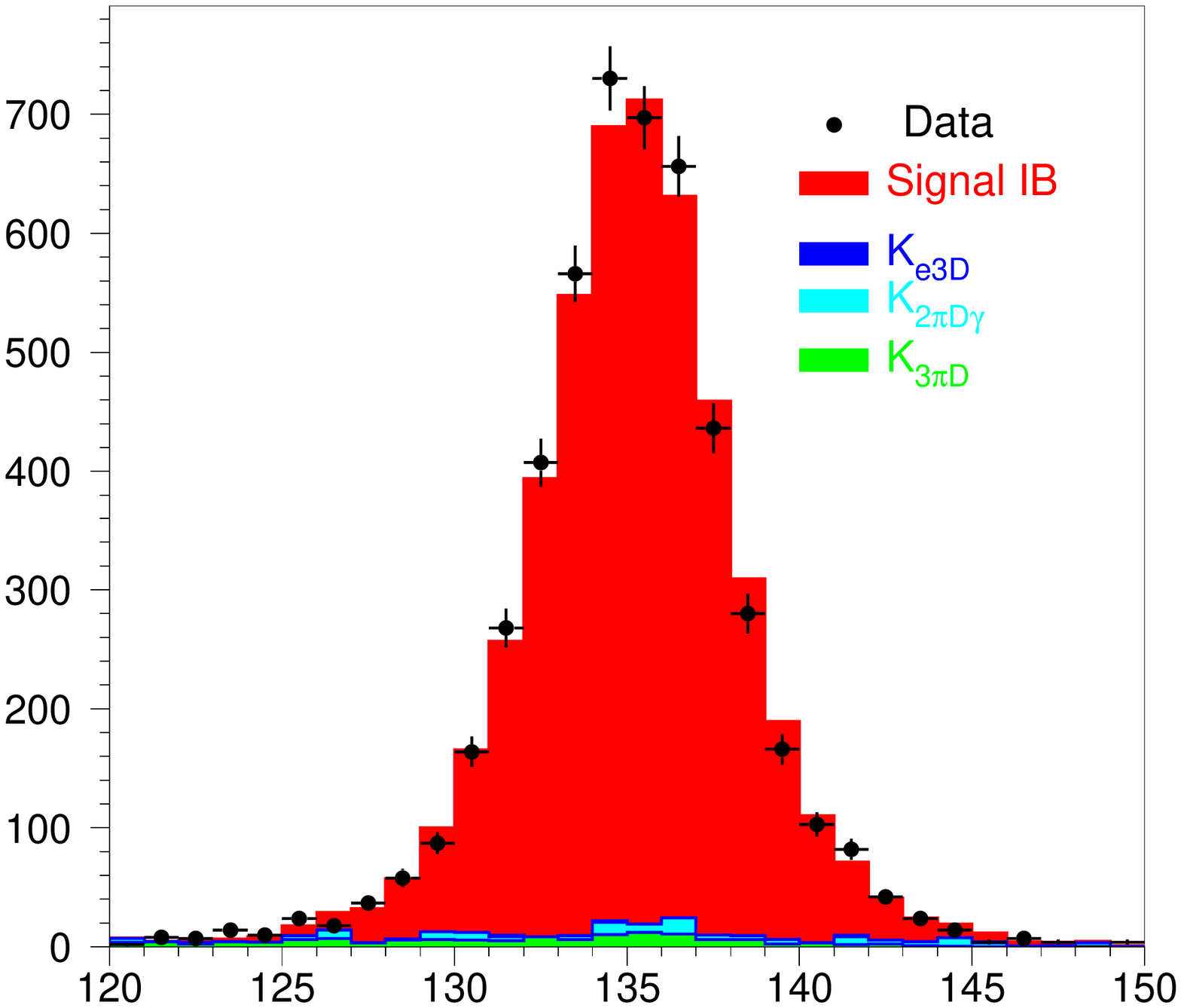}
\put(-90,-5){$m_{\gamma \gamma}$ [$\MEVcc$]}
\put(-240,90){\rotatebox{90}{Events / (1 $\MEVcc$)}}
\end{minipage} 
\begin{minipage}{0.5\linewidth}
\includegraphics[width=1.\linewidth]{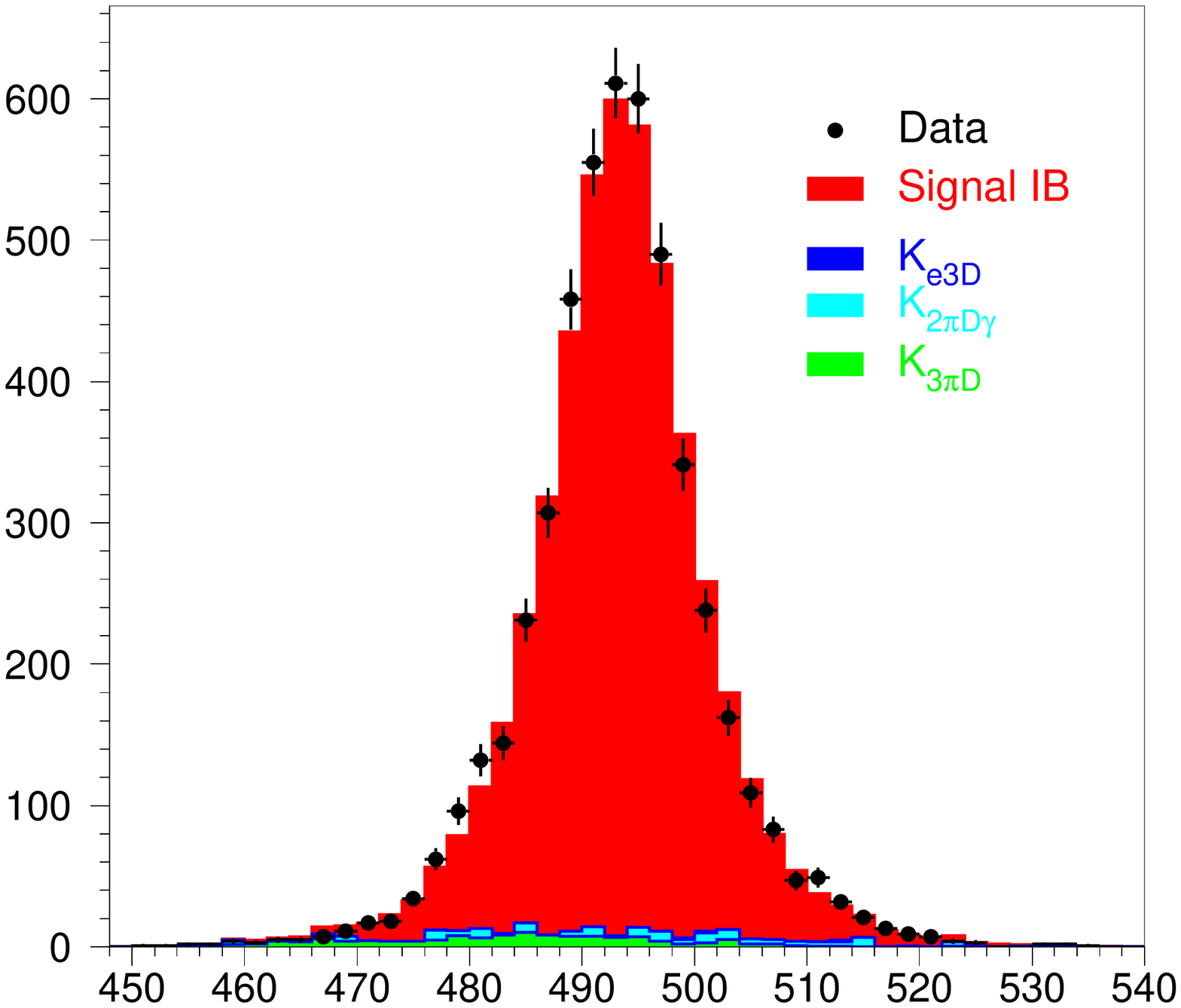}
\put(-110,-5){$m_{\pi^\pm \pi^0  e^+ e^-}$ [$\MEVcc$]}    
\put(-240,90){\rotatebox{90}{Events / (2 $\MEVcc$)}}
\end{minipage} 
\caption{ \label{fig:ppee} 
Signal candidates.
Left: reconstructed  $\gamma\gamma$ 
mass. Right: reconstructed  $\pi^\pm \pi^0 e^+ e^-$  mass. 
Full dots correspond to data candidates; stacked histograms are, from bottom to top, the expected $K_{3 \pi D}$ (green), $K_{2 \pi D\gamma}$ (light blue)  and $K_{e3 D}$ (dark blue) backgrounds  and IB signal (red) estimated from simulation. All quoted errors are statistical.}
\end{figure}

\paragraph{Candidates and background} Samples of $16.3\times 10^6 K_{2\pi D}$ candidates and 4919 signal candidates have been selected from a subset of a $1.7 \times 10^{11}$ kaon decay exposure in 2003--2004.  
The background estimates from simulation amount to ($10437\pm 119$)
$K_{\mu3D}$ events and ($6851 \pm 106$) $K_{e3D}$ events in the normalization mode, corresponding to a total  relative background contribution of 0.11\%.  In the signal  mode, 
they amount to ($132 \pm 8$) events from $K_{3\pi D}$,  ($102 \pm 19$) events from $K_{2\pi D\gamma}$ and  ($7 \pm 3$) from $K_{e3D}$,  adding up to a relative background contribution of (4.9 $\pm$ 0.4)\%.
 The reconstructed $\gamma e^+ e^-  ~(\gamma\gamma)$ 
 and $\pi^\pm \pi^{0}_D$ ($\PPEE$) mass distributions are displayed in \Fig{fig:pp0d} (\Fig{fig:ppee}) for the selected normalization (signal) candidates.  
 Background and normalization (signal) simulations, scaled to the number of observed candidates, show a good agreement with the data distributions. 
\paragraph{Acceptances} Because the selection acceptance is not uniform across the phase space, its overall value depends on the dynamics of the considered process. 
 The acceptance $A_n$  (3.981\%) is computed using the simulation of $K^\pm \to \pi^\pm \pi^0 $ according to \cite{gatti} followed by $\pi^0 _D$ decay according to the most recent ``Prague''  radiative decay calculation  \cite{Husek}. 
 
\begin{figure}[h]
\begin{minipage}{0.5\linewidth}
\includegraphics[width=1.\linewidth]{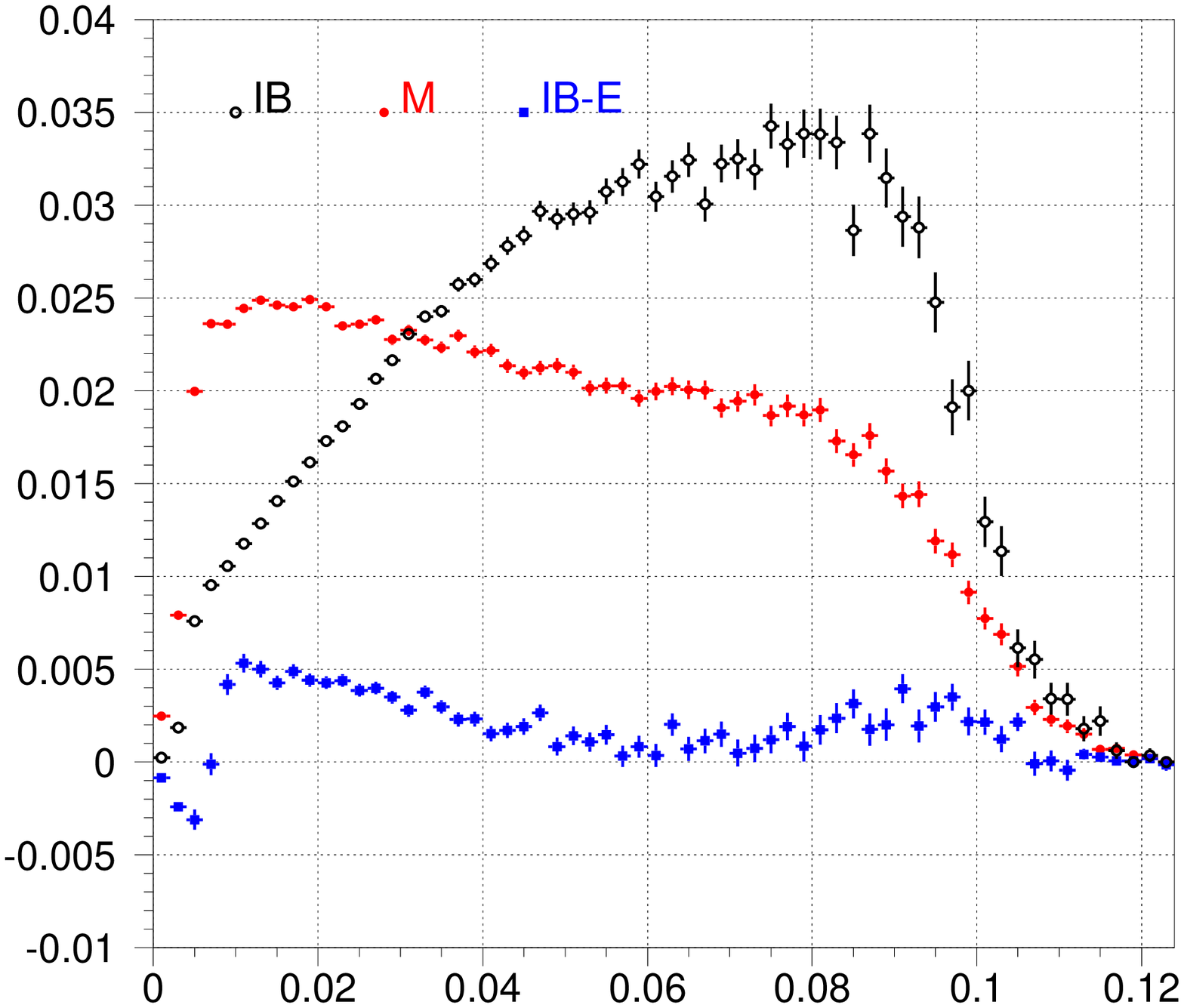}
\put(-95,-5){$\MEE$ [$\GEVcc$]}
\end{minipage}
\begin{minipage}{0.5\linewidth}
\includegraphics[width=1.\linewidth]{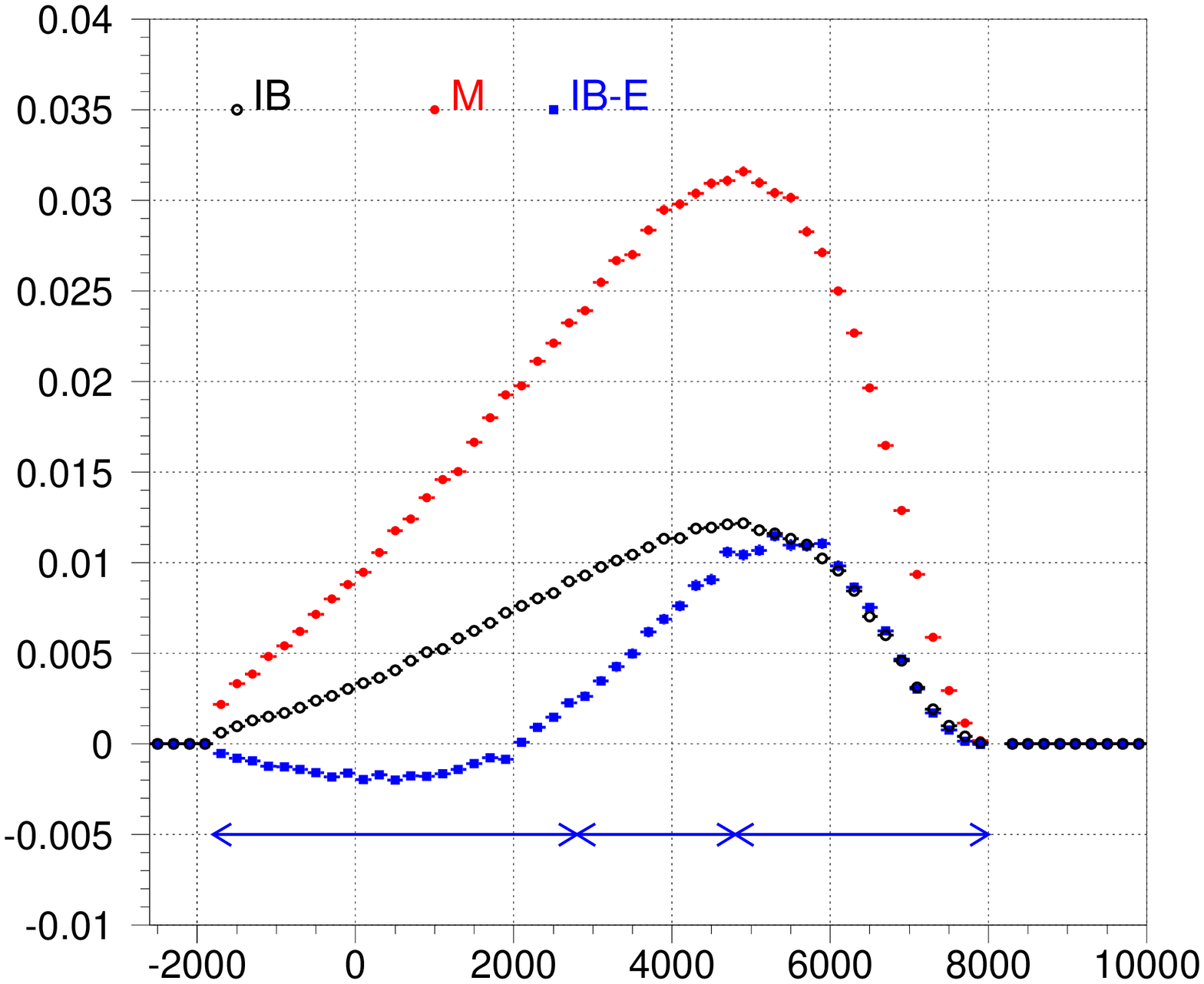}
\put(-95,-5){$Z_{vertex}$ [cm]}
\end{minipage}
\caption{ \label{fig:acc}   Acceptances of the IB, M and IB-E components  projected along the $\MEE$ and the longitudinal vertex position $Z_{vertex}$ variables (the  $Z$ axis origin is located 18 m downstream of the last collimator exit).  For the IB-E component, the acceptance is formally plotted  with a negative (positive) value when the interference is destructive (constructive). The arrows correspond to the three exclusive regions considered.}  
\end{figure}

The MC samples  for the different $\KPM \to \pi ^\pm \pi^0 e^+ e^-$  signal contributions IB, DE and INT have been generated separately according to the theoretical description given in \cite{Cappiello:2011qc,neweval}:
the DE contribution consists  mainly of the magnetic M term, with the  E term  expected to be fifteen times lower;
the INT term includes only the electric interference IB-E, as the other interference terms  IB-M and E-M do  not contribute to the total rate in the limit of full angular integration (\Sec{sec:intro}).
Particular care has been taken in the generation of the IB-E term which contributes constructively or destructively  to the differential rate
depending  on the kinematic space region considered. This property is illustrated  in \Fig{fig:acc}-left.
Radiative effects are implemented using the {\tt PHOTOS} package \cite{Photos}.

 Global acceptances are obtained for each of the three main components of the signal process: IB $(0.645 \pm 0.001)$\%, M $(1.723 \pm 0.003)$\% and  \text{IB-E} $(0.288 \pm 0.001)$\%. The signal acceptance $A_s$  is then obtained from a weighted average of the single-component acceptances, using as weights, $w$,  their relative contributions to the total rate with respect to IB 
computed in \cite{Cappiello:2011qc,neweval}: 
\begin{equation}
A_s=\frac{A_{\rm IB} + A_{\rm M} \cdot w_{\rm M} + A_{\text{IB-E}} \cdot w_{\text{IB-E}} } {1 + w_{\rm M} + w_{\text{IB-E}}} ,
\end{equation}
 where $w_{\rm M}$ and $w_{\text{IB-E}}$ are equal to 1/71 and  $-1/253$  respectively.
 The resulting signal acceptance is obtained as 
 $A_s = 0.9900 ~A_{\rm IB} +  0.0139 ~A_{\rm M} -0.0039 ~A_{\text{IB-E}} = (0.662 \pm 0.001)$\%.

Both normalization and signal acceptances are obtained with respect to the full $m_{ee}$ kinematic range.
\paragraph{Trigger efficiencies}
 Trigger efficiencies are measured from control data samples for the normalization mode (L1: $(99.75\pm 0.01)$\%, L2: $(97.66\pm 0.04)$\%) and cross-checked against the simulated estimations (L1: $(99.767\pm 0.003)$\%, L2: $(98.495\pm0.006)$\%) which provide also an accurate description of their time variations due to local and temporary inefficiencies of the HOD or DCHs.   Due to the low statistics of the signal candidate sample, it is not possible to obtain the trigger efficiencies from the downscaled control samples.   
 Trigger efficiencies for the signal candidates are  therefore estimated from the simulated samples  (L1:  $(99.729\pm0.009)$\%, L2:  $(98.604\pm0.021)$\%) and not affected by otherwise  large statistical uncertainties.  The full trigger efficiency in each selection is obtained as the product of L1 and L2 efficiencies that are based on different detectors and therefore uncorrelated.

 \paragraph{Systematic uncertainties}\label{sec:syst}
The statistical uncertainties on acceptance and trigger efficiency values are accounted as part of the systematic uncertainties.

The control of the geometrical acceptances is evaluated by considering three exclusive regions of the decay longitudinal position (shown in \Fig{fig:acc}-right)
with different acceptances and background conditions  
for both signal and normalization channels.
The difference between the statistical combination of the three $BR$ values and the global value is quoted as systematic uncertainty.

The  control  of the acceptance dependence with time and kaon charge  is  quantified by considering four exclusive  $BR$ measurements (2003 and 2004 data sets, $\KPL$ and $\KMI$ decays) and quoting as systematic uncertainty the difference between the statistical combination of the four $BR$ values and the global value.

An evaluation of the background control  level is obtained  by tightening the constraint of \Equ{eq:correl} to reduce the background to signal contribution from 4.9\% to 3\% while  decreasing the signal acceptance by  a relative fraction of 8\%.  
The quoted uncertainty covers also the effect of the residual disagreement between data and simulated reconstructed masses.

Trigger efficiencies obtained from simulation are used in the $BR$ calculation. The 
difference between the measured and simulated efficiencies  of the normalization candidates is considered as a systematic uncertainty. 

 The model dependence of the signal acceptance is investigated by varying in turn each  input ($N_{M}^{(0)},N_{E}^{(0,1,2)}$) within 
 its theoretical uncertainty estimate. The  resulting variations in acceptance are added in quadrature to 
 obtain the  overall  contribution to systematics.
 
\begin{figure}[h]
\begin{minipage}{0.5\linewidth}
\includegraphics[width=1.\linewidth]{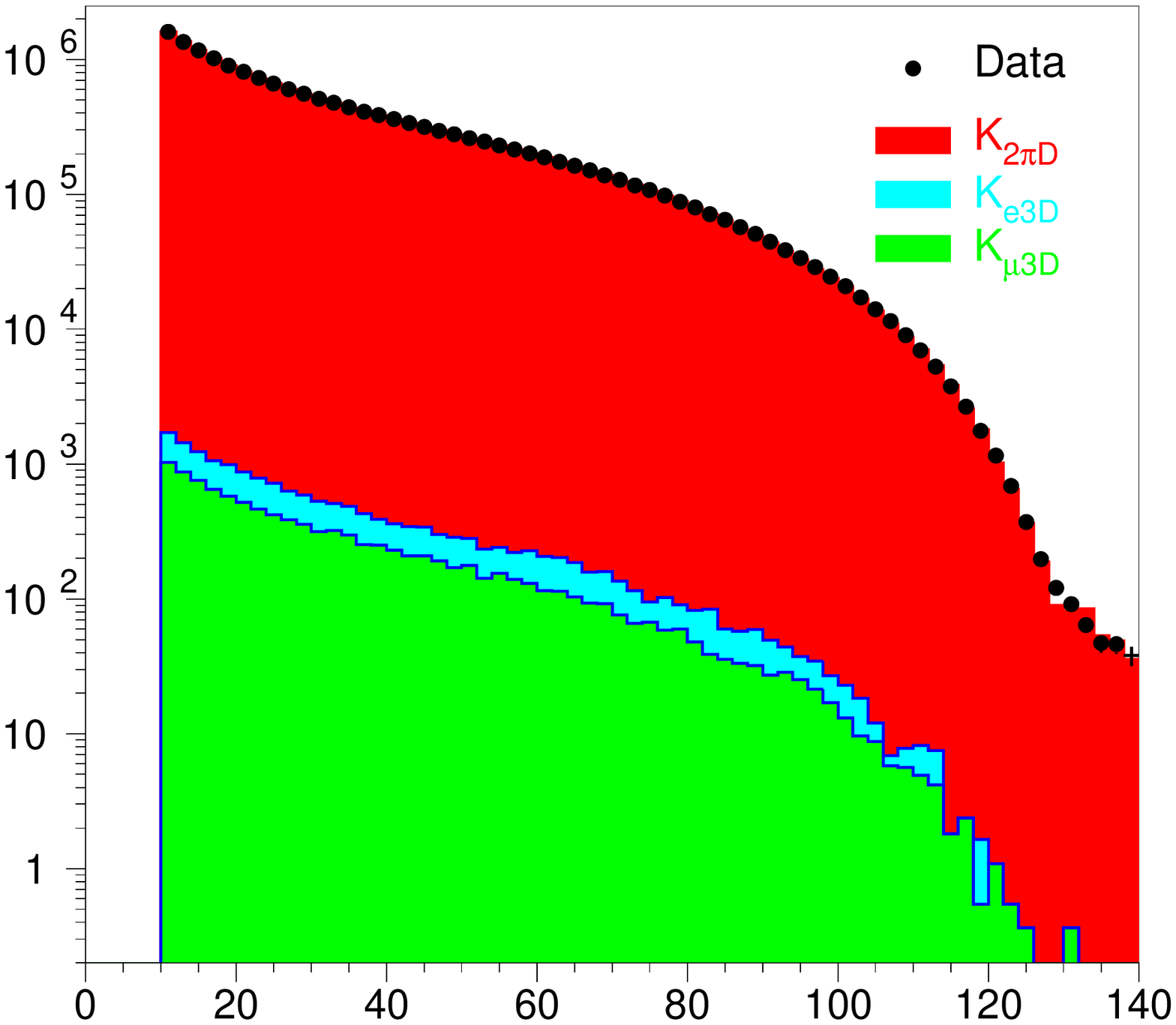}
\put(-90,-5){$\MEE$ [$\MEVcc$]}
\put(-130,175){$\PPD$}
\put(-240,90){\rotatebox{90}{Events / (2 $\MEVcc$)}}
\end{minipage}
\begin{minipage}{0.5\linewidth}
\includegraphics[width=1.\linewidth]{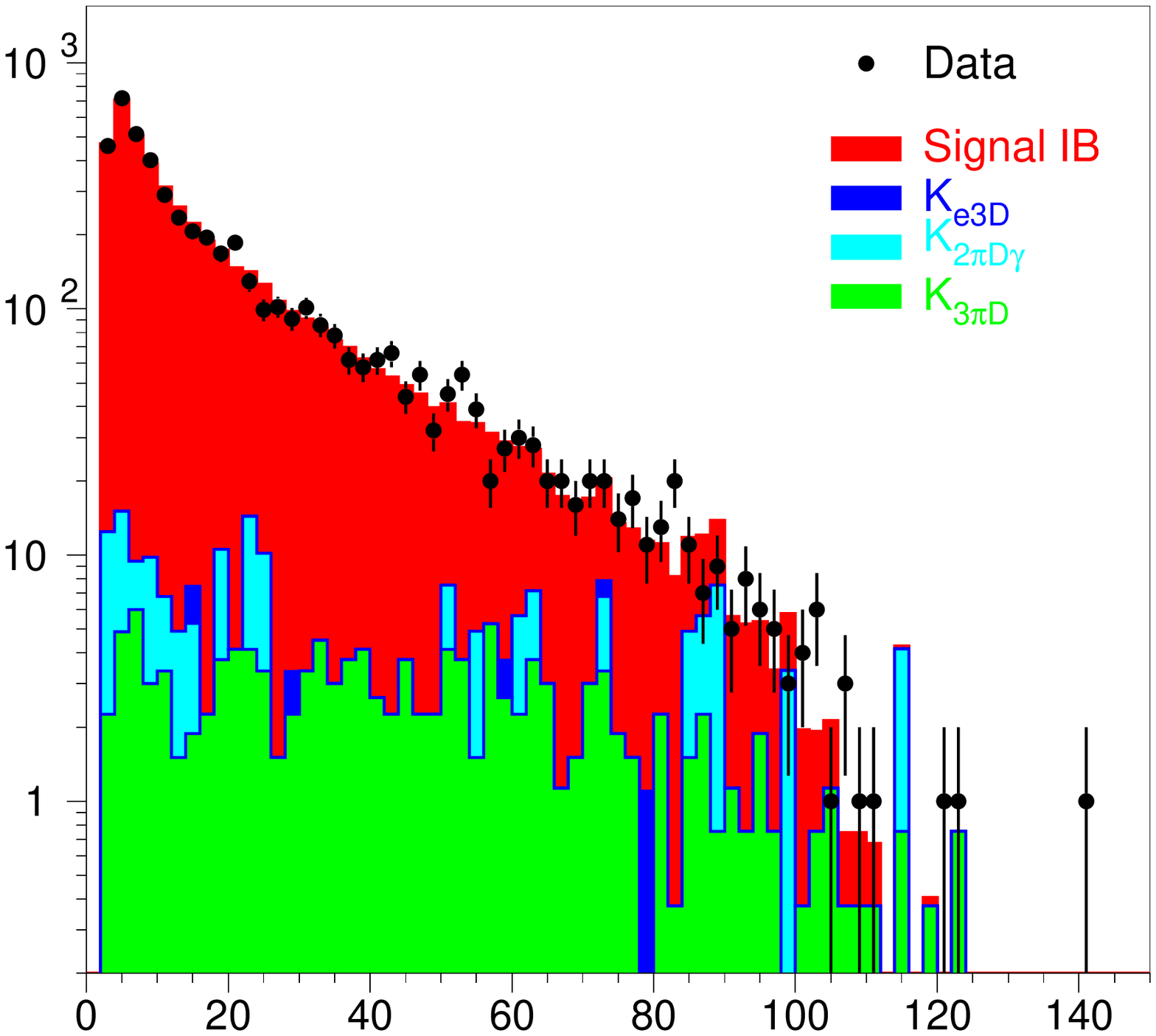}
\put(-95,-5){$\MEE$ [$\MEVcc$]}
\put(-130,175){$\PPEE$}
\put(-240,90){\rotatebox{90}{Events / (2 $\MEVcc$)}}
\end{minipage}
\caption{ \label{fig:mee}  Reconstructed $e^+  e^-$ mass distribution 
 for the normalization  (left) and   signal (right) candidates  with the lower cuts of 10 and 3~$\MEVcc$, respectively. Simulated background  and normalization (signal) contributions are also displayed.}
\end{figure}

 According to the 
 authors of the {\tt PHOTOS} package \cite{photos2}, the uncertainty on the photon emission implementation cannot exceed 10\% of 
 the full effect (here $4.9 \times 10^{-2}$ relative in the signal mode), which is quoted as systematic uncertainty. In the normalization mode,  
 in the absence of any prescription from the authors of the  
 ``Prague''  $\Pod$ decay implementation, 10\% of the 
 $0.53 \times10^{-2}$ relative difference between the {\tt PHOTOS}  and ``Prague''  
 $K_{2\pi D}$ acceptances
is conservatively assigned as a systematic uncertainty and added quadratically to the signal  
{\tt PHOTOS}  uncertainty. The agreement between data and simulation can be judged from the $\MEE$ distributions of  \Fig{fig:mee}.

External errors stem from  relative errors on $BR$($\KTP$) and  on 
$\Gamma$$(\pi^{0}_D)$/$\Gamma$$(\pi^{0}_{\gamma\gamma})$.

\Tab{tab:syst} summarizes the considered sources of uncertainty. 
 \begin{table}[htb]
\caption{\label{tab:syst}Statistical, systematic and external 
uncertainties to the $\KPPEE$ branching ratio measurement. The uncertainties related to
the model dependence and to radiative effects can also be 
considered as external errors as being unrelated to our data.}
\begin{center}
\begin{tabular}{l| c}
Source & $\delta BR/BR \times 10^{2}$\\ 
\hline
$N_s$ & 1.426\\
$N_{bs}$ & 0.416\\
$N_n$ & 0.025\\
$N_{bn}$ & negl.\\
\hline
{\bf Total statistical} & {\bf 1.486} \\
\hline
$A_s$ (MC statistics)& 0.171 \\
$A_n$ (MC statistics)& 0.051 \\
$\varepsilon({\rm L1}_s \times {\rm L2}_s$) (MC statistics)& 0.023\\ 
$\varepsilon({\rm L1}_n \times {\rm L2}_n$) (MC statistics)& 0.007\\ 
Acceptance geometry  control & 0.083\\
Acceptance time variation control & 0.064\\
Background control & 0.280 \\
Trigger efficiency (systematics)& 0.400 \\
Model dependence & 0.285 \\
Radiative effects & 0.490\\
\hline
{\bf Total systematic} & {\bf 0.777 }\\
\hline
$BR$($K_{2\pi}$) & 0.387 \\
$\Gamma (\pi^{0}_D) / \Gamma (\pi^{0}_{\gamma\gamma})$ &2.946 \\
\hline
{\bf Total external} & {\bf 2.971} \\
\end{tabular}
\end{center}
\end{table}

\paragraph{Result}
 The final result is obtained as:
\begin{equation}
BR (\KPPEE) = (4.237 \pm 0.063_{stat} \pm 0.033_{syst} \pm 0.126_{ext})\times ~10^{-6},
\end {equation}
 where the statistical error is dominated by the signal statistics, the systematic error by the radiative effects  and the external error by the $\Pod$ branching ratio uncertainty.

This value can be compared to the predictions from \cite{Cappiello:2011qc,neweval}:
 $BR(\KPPEE) = 4.183 \times 10^{-6}$ for IB only,
$ BR(\KPPEE) = 4.229 \times 10^{-6}$ when including all DE and INT terms.  
 The obtained value is compatible with both predictions within the experimental errors.
 However it should be noted that none of the above predictions includes any radiative  or isospin breaking 
effects.

\section{Kinematic space  study}
 The current data statistics does not allow  a precise enough measurement to quantify the contribution of the DE magnetic term M to the total decay rate (expected to be about 1\%). However, 
 the authors of \cite{Cappiello:2011qc,neweval}  have pointed out that the
contributions of IB, magnetic M, and interference IB-E  terms  have different distributions in 
the Dalitz plot (T$^{*}_{\pi}$, E$^{*}_{\gamma}$)  for different ranges of $q^2$ values, where  T$^{*}_{\pi}$, 
E$^{*}_{\gamma}$ and $q^2$ are the charged pion kinetic energy and  the virtual photon energy in the kaon rest frame,  and the $e^+ e^-$  mass squared,  respectively.
 The differences remain  relevant even after the analysis selection acceptance is applied.
 A method based on the population of 3d-boxes in the kinematic space 
 ($q^2$, T$^{*}_{\pi}$, E$^{*}_{\gamma}$) is used to determine the relative fraction of each component  that would add up to reproduce the data sample   population. The data 3d-space  is first split into 
N1 slices along $q^2$, then each slice is split into N2 
slices along T$^{*}_{\pi}$ and then into N3 E$^{*}_{\gamma}$  slices, all with equal populations. The 
result is a grid of N1 $\times$ N2 $\times$ N3 exclusive 3d-boxes of variable 
size but identical population.  The background contributions and the various simulated signal components 
are distributed according to the data grid definition, each resulting in  a set of 3d-boxes of unequal  
population.  To account for the potentially different sizes of the simulation samples, scale factors $\rho_{\rm M}$ and $\rho_{\text{IB-E}}$ are defined as the ratios of the IB  to the M and  IB to the IB-E simulated sample sizes.

To obtain the fractions (M)/IB and (IB-E)/IB reproducing the data, a $\chi^2$ estimator is minimized:
\begin{equation}\label{eq:chi2}
 \chi^2 = \sum_{i=1}^{{\rm N1} \times {\rm N2} \times {\rm N3}} (N_i - M_i )^2 /(\delta N_i ^2 + \delta M_i ^2 ),
\end{equation}
where $N_i  ~(\delta N_i )$ is the data population (error) and  $M_i  ~(\delta M _i )$ the expected population 
(error) in box $i$. The denominator of each term is dominated by the number of data events $\delta N_i ^2 = N_i$, the 
same in each box.
  The expected number of events in box $i$ is computed as:

\begin{equation}
 M_i = N \times ( N_i ^{\rm IB} + a \cdot  N_i ^{\rm M} + b \cdot  N_i ^{\text{IB-E}} ) + N_i ^{Bkg},
\end{equation}
 where $N$ is a global scale factor to guarantee that the sum of the simulated 
events and background contributions is normalized to the total number of data candidates.
 At the end of the minimization, the obtained values of $a$ and $b$ can be related to the 
relative  contributions (M)/IB and (IB-E)/IB by :

\begin{equation}
 (\text{M)/IB} = (a \pm \delta a)/ \rho_{\text {M}}, ~~~~~~~\text{(IB-E)/IB} = (b \pm \delta b)/ \rho_{\text{IB-E}}.
\end{equation}
The method  has no sizeable dependence on the precise grid structure as long as the granularity ensures sensitivity to the population variation within the resolution (at least 3 $q^2$ slices and 5 or 6  slices along the two other variables)  and large enough statistics per box  to consider Gaussian errors.
The grid configuration $3\times5\times6$  has been employed and the  results are obtained with a 
$\chi^2$ probability of 
19\% for a value of 98.2/87 degrees of freedom and a correlation $C(a,b) = 0.06$.  
The obtained value (M)/IB = $0.0114 \pm 0.0043_{stat}$ is consistent with 
the predicted value from \cite{Cappiello:2011qc}, $1/71 = 0.0141 \pm 0.0014_{ext}$, obtained using 
the experimental measurement of $N^{(0)}_M $.
The (IB-E)/IB value of $-0.0014 \pm 0.0036_{stat}$ shows that there is no sensitivity
to this contribution within the current data statistics  
and agrees with the value from \cite{neweval}, $-1/253 = -0.0039 \pm 0.0028_{ext}$,
obtained using experimental inputs to $N^{(0,1,2)}_E$ values.
The external errors on the predicted values stem from the  uncertainties of the measurements used as input in the evaluations.

\section{Asymmetry investigations}\label{sec:asym}
 Electroweak (or beyond Standard Model) phases change sign 
 under charge conjugation when switching from $\KPL$ to $\KMI$, unlike the strong phase $\delta = \delta_0 ^2 - \delta_1 ^1$ that governs the final state interaction of the pion system. These phases can be investigated through asymmetries between $\KPL$ and $\KMI$ partial rates.

The simplest CP-violating  asymmetry 
 is the charge asymmetry between $\KPL$ and $\KMI$ partial rates integrated over the whole phase space:
\begin{equation}
A_{CP} =  \frac{\Gamma (\PPEEP) - \Gamma (\PPEEM) }{\Gamma (\PPEEP) + \Gamma (\PPEEM)}.
\end{equation}
 The value of  $A_{CP} $ can be related to the interference IB-E term and is proportional to $\sin\delta \sin\Phi_E$, where 
 $\Phi_E$ is a possible CP-violating phase appearing in the form factors $F_1 ^{DE},$ $F_2 ^{DE}$  in addition (subtraction)  to the strong phase $\delta_1 ^1$
 (\Sec{sec:intro}). The asymmetry is obtained from the statistically independent measurements of 
 $\KPL$ and $\KMI$ branching ratios,  that take into account the  possible biases introduced by the detector acceptances.
 The values 
 \begin{equation}
 BR(\KPL) = (4.151 \pm 0.078_{stat}) \times 10^{-6}, ~BR(\KMI) = (4.394 \pm 0.108_{stat}) \times 10^{-6}
 \end{equation}
lead to $A_{CP} = -0.0284 \pm 0.0155$,
 where the error is statistical only, as the systematic and external errors cancel in the ratio.
This value is consistent with zero 
and is translated to a single-sided limit: 
\begin{equation}
| A_{CP}|< 4.82 \times 10^{-2}  {\rm  ~at~90\%~CL.}
 \end{equation}
 
Other asymmetries are defined in \cite{Cappiello:2011qc} using the so-called Cabibbo-Maksymowicz  \cite{cabibbo65} variables\footnote{For $\KPM$ decays, the variables are the squared invariant dipion and dilepton masses,  
the angle of the $\pi^{\pm}$ ($e^{\pm}$) in the dipion (dilepton) rest frame with respect to the flight direction of the dipion (dilepton) in the $\KPM$ rest frame,  the angle $\phi$ between the dipion and dilepton planes in the kaon rest frame.}  
to describe the kinematic space of the decay and selecting particular integration regions of the $\phi$ angular variable: 

\begin{equation}\label{eq:star}
 A_{CP}^{\phi^*} =\frac{\displaystyle\int_0^{2\pi}\frac{d\Gamma_{(K^+ -K^-)}}{d\phi}d\phi^*}{\displaystyle\int_0^{2\pi}\frac{d\Gamma_{(K^+ + K^-)}}{d\phi}d\phi},
{\rm ~where~} 
\int_0^{2\pi}d\phi^*\equiv \left[\int_0^{\pi/2}-\int_{\pi/2}^{\pi}+\int_{\pi}^{3\pi/2}-\int_{3\pi/2}^{2\pi}\right]d\phi ,\end{equation}
 
\begin{equation}\label{eq:tilde}
A_{CP}^{\tilde{\phi}} =\frac{\displaystyle\int_0^{2\pi}\frac{d\Gamma_{(K^+ - K^-)}}{d\phi}d{\tilde{\phi}}}{\displaystyle\int_0^{2\pi}\frac{d\Gamma_{(K^+ + K^-)}}{d\phi}d\phi},
{\rm ~where~} 
 \int_0^{2\pi}d{\tilde{\phi}}\equiv \left[\int_0^{\pi/2}+\int_{\pi/2}^{\pi}-\int_{\pi}^{3\pi/2}-\int_{3\pi/2}^{2\pi}\right]d\phi .
 \end{equation}
These asymmetries can be obtained by combining the branching ratios measured in
 various parts of the $\phi$ variable space. Defining sectors of the 
$\phi$  space between 0 and $2\pi$ as $\Phi1 ~(0, \pi /2)$,  
$\Phi2 ~(\pi /2, \pi)$, $\Phi3 ~(\pi, 3\pi /2)$ and
 $\Phi4$ $(3\pi /2 , 2\pi)$,  and combining them as statistically 
independent sector sums ($\Phi13 = \Phi1 + \Phi3, ~\Phi24= \Phi2 + \Phi4$) and ($\Phi12 = \Phi1 + \Phi2, ~\Phi34 = \Phi3 + \Phi4$) one can 
obtain the above asymmetries. 

The $\phi^*$ integral has the interesting property of subtracting the 
contribution of sector sum $\Phi24$ from the contribution of sector sum $\Phi13$. The 
interference term IB-M (Section \ref{sec:intro}) equally populates sectors $\Phi1$
 and $\Phi3$ when positive and depopulates sectors $\Phi2$ and $\Phi4$ when negative.
The $ A_{CP}^{\phi^*}$ asymmetry is then related to the interference IB-M term and is 
proportional to $\cos\delta \sin\Phi_M$, where $\Phi_M$ is a possible CP-violating phase
appearing in the form factor $F_3 ^{DE}$ (Section \ref{sec:intro}). 
The interference IB-M term has not been generated in the simulation as it is not expected to contribute significantly 
to the total rate. However it has been checked that the whole 
range of the $\phi$ variable is always considered in the acceptance 
calculation, apart for the region $q^2 < 3$ (MeV/$c^2)^2$ excluded from the 
signal selection.
The CP asymmetries defined in Eq.~(\ref{eq:star}, \ref{eq:tilde}) 
are measured, although to a limited precision given the current data statistics, as:
\begin{equation}
A_{CP}^{\phi^*} =   0.0119 \pm 0.0150_{stat}  {\rm ~~~and~~~ } A_{CP}^{\tilde{\phi}} =   0.0058 \pm 0.0150_{stat}.
 \end{equation}
All asymmetries are consistent with zero,  single-sided upper limits can be 
set as
 
\begin{equation}
 |A_{CP}^{\phi^\ast} | < 3.11 \times 10^{-2} ,  |A_{CP}^{\tilde{\phi}} | < 2.50 \times 10^{-2} {\rm ~at~90\%~CL.}
 \end{equation}
Following another prescription of \cite{Cappiello:2011qc}, a long-distance P-violating  asymmetry 
defined as
\begin{equation}
A_{P}^{(L)} = \frac{\displaystyle\int_0^{2\pi}\frac{d\Gamma}{d\phi}d\phi^*}{\displaystyle\int_0^{2\pi}\frac{d\Gamma}{d\phi}d\phi} = \frac{\Gamma(\Phi13) - \Gamma(\Phi24)}{\Gamma}
 \end{equation}
 can be obtained from the asymmetry  between sector sums $\Phi13$ and $\Phi24$  when considering 
 $\KPL$ or $\KMI$ alone, and combined if found consistent.
The $A_{P}^{(L)}$ asymmetry is proportional to $N_M ^{(0)}$ \cite{Cappiello:2011qc} and $\sin\delta$.
A precise $A_{P}^{(L)}$ measurement would allow a check of the sign of 
$N_M ^{(0)}$ and a measurement of $\sin\delta$.

Our data lead to  
$A_{P}^{(L)} (\KPL) = 0.0059 \pm 0.0180_{stat}$ and $A_{P}^{(L)} (\KMI) = -0.0166 \pm 0.0237_{stat},$
both consistent with zero.
The combined value is $A_{P}^{(L)} (\KPM) = -0.0023 \pm 0.0144_{stat} $. The errors
 are statistical only as both systematic and external uncertainties cancel in
 the ratios. This value can be translated into a single-sided upper limit:

\begin{equation}
| A_{P}^{(L)}| < 2.07 \times 10^{-2} {\rm ~at~90\%~CL.}
 \end{equation}

\section{Results and conclusion}
The data sample recorded by the NA48/2 experiment in 2003--2004 has been analyzed, searching 
for the unobserved $\KPPEE$ decay mode in an exposure of $1.7 \times 10^{11}$ kaon decays.
 A sample of 4919 decay candidates with 4.9\% background has been identified, resulting in
  the  first observation of this decay mode. The branching ratio has been measured relative 
to the $\KDA$ mode followed by a Dalitz decay $\PDAL$ and found to be 
$(4.237 \pm 0.063_{stat} \pm 0.033_{syst} \pm 0.126_{ext} ) \times 10^{-6}$, 
in agreement with predictions from ChPT.  

Despite the limited statistics available, a study of the kinematic 
space of the decay has been performed to extract information on the fraction of  magnetic (M)
and interference (IB-E) contributions with respect to inner bremsstrahlung (IB). The relative 
contribution, (M)/IB = $(1.14 \pm 0.43_{stat}) \times 10^{-2}$, is found consistent with 
the theoretical expectation of $(1.41\pm 0.14_{ext}) \times 10^{-2}$. The 
relative IB-E contribution, (IB-E)/IB = $(-0.14 \pm 0.36_{stat}) \times 10^{-2}$, is 
also in agreement with the prediction of $(-0.39 \pm 0.28_{ext}) \times 10^{-2} $ 
but with limited significance due to the lack of data statistics in the high 
$\MEE$ region. 

Several CP-violating asymmetries and a long-distance P-violating asymmetry have
been evaluated and found to be consistent with zero, leading to  
upper limits 
$|A_{CP}| < 4.8 \times 10^{-2}, ~|A_{CP}^{\phi^\ast} | < 3.1 \times 10^{-2},~|A_{CP}^{\tilde\phi} | < 2.5 \times 10^{-2}, ~|A_{P}^{(L)}| < 2.1 \times 10^{-2}$ at 90\% CL.

 %
If larger data statistics becomes available (for example at the  NA62 experiment), more 
detailed studies of the kinematic space will allow for an improved
evaluation of the DE term contribution.  
A study of the  P-violating asymmetry 
could bring information on the sign of the DE magnetic term  and on the strong phase $\delta$ involved in the final state interaction of the two pions.

\section*{Acknowledgements}
We gratefully acknowledge the CERN SPS accelerator and beam line staff for the excellent
performance of the beam and the technical staff of the participating institutes for their efforts
in the maintenance and operation of the detector, and data processing. We thank M. Koval for making the  ``Prague'' radiative $\PIo$ Dalitz decay code available in the NA48/2 simulation software. Discussions with G.~D'Ambrosio and O.~Cat\`a were most stimulating in clarifying the impact of interference terms on our measurement. 



\begin{thebibliography}{99}
\bibitem{Cirigliano:2011ny}
  V.~Cirigliano, G.~Ecker, H.~Neufeld, A.~Pich and J.~Portol\'es,
  Rev.\ Mod.\ Phys.\  {\bf 84} (2012) 399.
%
\bibitem{Pichl:2000ab} 
  H.~Pichl,  Eur.\ Phys.\ J.\ C {\bf 20} (2001) 371.
%
\bibitem{Cappiello:2011qc} 
  L.~Cappiello, O.~Cat\`a, G.~D'Ambrosio and  D.N.~Gao, Eur.\ Phys.\ J.\ C {\bf 72} (2012) 1872.
%
\bibitem{Gevorkyan:2014waa} 
  S.R.~Gevorkyan and M.H.~Misheva,   Eur.\ Phys.\ J.\ C {\bf 74} (2014) 2860.
%
\bibitem{Batley:2010aa} 
J.R.~Batley {\it et al.}  [NA48/2 Collaboration], Eur.\ Phys.\ J.\ C {\bf 68} (2010) 75.
%
\bibitem{neweval}
  L.~Cappiello, O.~Cat\`a and G.~D'Ambrosio,
Eur.\ Phys.\ J.\ C {\bf 78} (2018) 265.
%
 \bibitem{Batley:2007} 
 J.R.~Batley  {\it et al.}  [NA48/2 Collaboration],  Eur.\ Phys.\ J.\ C {\bf 52} (2007) 875.
%
\bibitem{fa07}
V.~Fanti  {\it et al.}  [NA48 Collaboration],  Nucl. Instrum. Methods A {\bf 574} (2007) 443.
%
\bibitem{pdg}
M.~Tanabashi  {\em et al.}  (Particle Data Group), Phys. Rev. D {\bf 98} (2018) 030001.
%
\bibitem{GEANT}
GEANT3  Detector Description \& Simulation Tool,  CERN Program Library {\bf W5013} (1994).
%
\bibitem{gatti}
C.~Gatti, Eur.\ Phys.\ J.\ C {\bf 45} (2006) 417.
%
\bibitem{Husek} 
T.~Husek, K.~Kampf and J.~Novotn\'y,  Phys.\ Rev.\ D {\bf 92}  (2015) 054027.
%
\bibitem{Photos} 
E.~Barberio and Z.~W\c{a}s, Comput. Phys. Commun. {\bf 79} (1994) 291.
%
\bibitem{photos2} Qingjun~Xu, Z.~W\c{a}s, Proceedings of Science, PoS~(RADCOR2009)~(2009) {\bf 071}.
%
 \bibitem{cabibbo65} 
N.~Cabibbo and A.~Maksymowicz, Phys.  Rev. {\bf 137}  (1965) B438.

\end{thebibliography}
\end{document}